\documentclass[sigconf]{acmart}

\AtBeginDocument{%
  \providecommand\BibTeX{{%
    \normalfont B\kern-0.5em{\scshape i\kern-0.25em b}\kern-0.8em\TeX}}}

\setcopyright{acmcopyright}
\copyrightyear{2026}
\acmYear{2026}
\acmDOI{XXXXXXX.XXXXXXX}

\acmConference[KDD '26]{Proceedings of the 32nd ACM SIGKDD Conference on Knowledge Discovery and Data Mining}{August 2026}{Jeju, Korea}
\acmISBN{978-1-4503-XXXX-X/26/08}

\RequirePackage{amsmath}
\RequirePackage{amsthm}
\RequirePackage{nicefrac}

\usepackage{bbm,graphicx,bm}
\usepackage{multirow}
\usepackage{booktabs}
\usepackage{subcaption}
\usepackage{fontawesome5}
\usepackage{tabularx}
\newcommand{\robot}{\text{\textcolor{gray}{\scalebox{0.9}{\faRobot}}}}
\newcommand{\robotbar}{\mkern 4mu\overline{\mkern-2mu\robot\mkern-2mu}\mkern 4mu}
\newcommand{\robothat}{\mkern 4mu\widehat{\mkern-2mu\robot\mkern-2mu}\mkern 4mu}

\theoremstyle{definition}

\newcommand{\E}{\ensuremath{\mathbb{E}}}

\newcommand{\Cov}{\text{Cov}}
\newcommand{\Var}{\text{Var}}

\def\b1{\boldsymbol{1}}

\definecolor{RED}{RGB}{255,0,0}

\usepackage{soul}
\usepackage[utf8]{inputenc}

\title{AI-Assisted Variance Reduction in Randomized Experiments}

\author{David Arbour}
\affiliation{%
  \institution{Adobe Research}
  \city{San Jose}
  \state{CA}
  \country{USA}
}

\author{Eli Ben-Michael}
\affiliation{%
  \institution{Carnegie Mellon University}
  \city{Pittsburgh}
  \state{PA}
  \country{USA}
}

\author{Avi Feller}
\affiliation{%
  \institution{University of California, Berkeley}
  \city{Berkeley}
  \state{CA}
  \country{USA}
}

\author{Apoorva Lal}
\affiliation{%
  \institution{OpenAI}
  \city{San Francisco}
  \state{CA}
  \country{USA}
}

\author{Lo-Hua Yuan}
\affiliation{%
  \institution{Airbnb}
  \city{San Francisco}
  \state{CA}
  \country{USA}
}

\begin{abstract}
Generative AI and large language models can produce realistic predictions of human behavior from rich, unstructured inputs with little to no task-specific training data. Recent work uses these ``digital twin'' predictions to supplement human responses in surveys and experiments.
We study the special case of using AI-generated predictions to reduce variance in randomized experiments. We argue that doing so requires no new estimators and that researchers can simply include AI predictions as covariates in standard regression adjustment, analogous to adjusting for a prognostic score.
A benefit of this approach is a ``do no harm'' property whereby the adjusted estimator reverts to the unadjusted difference in means when predictions are uninformative.
Other methods, such as variants of prediction-powered inference, do not have this guarantee.
We provide implementation guidance, including how to obtain continuous scores from discrete LLM outputs and how to use LLMs to featurize unstructured inputs as auxiliary covariates. We demonstrate these ideas in simulations and three empirical applications: a survey mega-study, an email marketing A/B test, and a large-scale technology platform experiment.
Overall, efficiency gains are real if modest, with greater benefits in studies that contain substantial text and other unstructured data. We also confirm the do no harm property empirically. Given these gains and limited costs, we recommend adjusting for AI-generated predictions as a regular empirical practice.
\end{abstract}


\begin{document}

\maketitle

\section{Introduction}\label{sec:intro}

The rise of large language models (LLMs) has renewed interest in ``silicon sampling,'' ``digital twins,'' and related frameworks that use AI to simulate or predict human responses in surveys and experiments \citep{argyle2023out, horton2023large, brand2023using}. While these approaches are fast and scalable, they are often poor substitutes for human respondents \citep{cummins2025threat,peng2025digital}. This has motivated a growing methodological literature on \emph{hybrid} estimators that combine AI predictions with ground-truth human data, for example, via prediction-powered inference \citep[PPI; ][]{angelopoulos2023prediction} or semiparametrically efficient estimation \citep{ruan2025calm}.

This paper focuses on randomized experiments, where AI predictions can be used purely for variance reduction. Our central message is that no new methods are needed: researchers can include AI predictions as covariates in standard regression adjustment, analogous to adjusting for the estimated prognostic score \citep{hansen2008prognostic,schuler2022increasing}. 
In fact, we can view regression adjustment as the (linearly) calibrated form of several recently proposed bias-corrected estimators \citep{vdL2026calibeating}.

Relative to the unadjusted estimator, regression adjustment ``does no harm,'' in the sense that it retains unbiasedness and (weakly) reduces variance \citep{lin2013agnostic}. When predictions are uninformative, the fitted coefficient shrinks toward zero and the estimator reverts to the difference-in-means; when they are informative, precision improves. Uncalibrated approaches, such as the model-assisted estimator \citep{sarndal2003model} and variants of PPI \citep{angelopoulos2023prediction}, lack this guarantee and can \emph{inflate} variance when prediction quality is modest. Regression adjustment also accommodates baseline covariates alongside AI predictions.

How much AI predictions improve precision depends directly on their quality.
Pre-trained foundation models tend to be more predictive than traditional tabular methods when the signal comes from complex or unstructured inputs that are otherwise difficult to featurize.
With already-structured tabular data, ridge regression and XGBoost can match or outperform LLMs alone.
Even here, adjusting for both AI predictions and tabular features gives the best of both worlds.
More broadly, LLMs can also \emph{featurize} unstructured inputs, yielding additional covariates that further improve precision.

We also discuss practical implementation issues that arise with binary and count outcomes. In these cases, we argue that researchers should adjust for continuous predicted probabilities rather than discretized predictions. We give practical strategies for obtaining such scores, including extracting log-probabilities from the output token distribution and averaging over multiple stochastic draws.

We demonstrate these ideas in synthetic experiments and three empirical applications: a mega-study of survey experiments with digital twins, an email marketing A/B test with tabular covariates, and a large-scale technology platform experiment with sequential user data. Across these settings, incorporating AI predictions via regression adjustment yields real if modest gains beyond traditional covariates. By contrast, less careful uses of AI predictions can lead to bias or inflated standard errors.

\subsection{Previous literature}\label{previous-literature}

\paragraph{Synthetic surveys and digital twins.}
A growing literature explores using LLMs to generate ``silicon samples'' as substitutes for human participants \citep{argyle2023out, horton2023large, brand2023using}. While promising for rapid prototyping, synthetic samples often exhibit substantial bias \citep{cummins2025threat,peng2025digital}, hindering their use in real-world deployments.
Recognizing these limits, several recent methods instead focus on combining AI predictions with ground-truth human responses. These methods divide into two broad camps.

  \paragraph{Hybrid estimation: finite-sample approaches.} Motivated by survey calibration and control variates, several papers combine randomization and human responses to guarantee exact (or near-exact) unbiasedness without relying on model correctness \citep{egami2023using}. Prediction-powered inference \citep[PPI;][]{angelopoulos2023prediction} has emerged as a particularly popular framework, with many extensions \citep{song2026demystifying}.
  As we discuss below, regression adjustment in randomized experiments is a special case \citep{lin2013agnostic,deng2023augmentation}.  
  \citet{vdL2026calibeating} explore calibrated forms of PPI and connect that framework to the semiparametric causal inference literature, including documenting the equivalence between linear calibrated PPI++ and linear regression. 
  Separately, \citet{bashari2025statistical} propose using conformal methods for finite-sample guarantees when combining human and silicon samples.

  \paragraph{Hybrid estimation: asymptotic and semiparametric approaches.}
  Motivated by semiparametric efficiency theory, several papers instead propose AIPW-style estimators that optimally combine AI predictions with real data \citep{ruan2025calm, bartolomeis2025efficient, engh2025llm}. See also \citet{byun2025valid}, who propose a flexible strategy based on generalized method of moments. These approaches are more general than the finite-sample approaches above, allowing for observational data and weaker identification assumptions. However, they typically require stronger regularity conditions and more complex estimation procedures. Importantly, in randomized experiments these estimators are asymptotically equivalent to the methods we discuss below.

\paragraph{Propagating uncertainty from AI-generated variables.} A separate line of work in econometrics addresses inference when AI- or ML-generated variables are used as regressors or outcomes in observational analyses, where prediction error can bias estimates and invalidate standard errors \citep{battaglia2024inference,ludwig2024large}. These concerns do not arise in our setting: under randomization, AI predictions enter only as auxiliary covariates for variance reduction, and the resulting ATE estimators are unbiased (model-assisted) or consistent (regression-adjusted) regardless of prediction quality. The black-box predictor's error is absorbed into the residual variance rather than the point estimate, so explicitly modeling prediction uncertainty is unnecessary for valid inference on the ATE.

\paragraph{Complementary uses of AI in experiments.} There is a growing literature on using AI to improve other aspects of experimental design and analysis. For example, \citet{gui2025leveraging} introduce generative stratification, using LLMs to synthesize covariate information for stratified randomization. \citet{saenger2024autopersuade} use LLMs to optimize persuasive interventions.
\citet{broska2025mixed} consider hybrid designs that combine human and synthetic samples.

\paragraph{Empirical studies of variance reduction in randomized experiments.} We add to a substantial literature in the biomedical and social sciences on empirically assessing the plausible variance reduction from covariate adjustment in randomized trials. 
\citet{schochet2010regression} reviews eight large-scale social science experiments, finding that covariate adjustment reduces variance by a median of roughly 25\%, ranging from under 10\% to over 50\% depending on the outcome.
\citet{shao2025benchmarking} benchmark 18 covariate-adjustment strategies across 50 biomedical RCTs, finding a median variance reduction of 13\% for continuous outcomes and 5\% for binary outcomes; notably, ML-based methods do not improve over simple linear regression in their settings.

\section{Problem setup}\label{problem-setup}

We consider a standard finite population framework for randomized experiments \citep{freedman2008regression, lin2013agnostic} with $n$ units indexed by $i = 1, \ldots, n$. We refer to these $n$ units interchangeably as the ``finite population'' and ``sample''. Each unit has potential outcomes $Y_i(1)$ and $Y_i(0)$, representing the outcomes under treatment and control, respectively. In a completely randomized experiment, we randomly assign $n_1$ units to treatment and $n_0 = n - n_1$ units to control, where the treatment assignment is denoted by $Z_i \in \{0, 1\}$; our setup immediately extends to other randomized designs. The observed outcome for unit $i$ is $Y_i^{\text{obs}}  = Y_i(0) + Z_i (Y_i(1) - Y_i(0))$. Our estimand of interest is the sample average treatment effect (SATE),
\[
  \tau \;=\; \bar{Y}(1) - \bar{Y}(0) \;=\; \frac{1}{n} \sum_{i=1}^n Y_i(1) - \frac{1}{n} \sum_{i=1}^n Y_i(0),
\]
where $\bar{Y}(z) = \frac{1}{n} \sum_{i=1}^n Y_i(z)$ denotes the mean outcome under treatment $z$.

We have access to baseline covariates $X_i \in \mathcal{X}$.
Importantly, we also have access to predictions from a black-box model, which we denote as $\robot_i(z)$ for unit $i$ under treatment $z \in \{0, 1\}$. Under the finite-population framework, both potential outcomes $Y_i(z)$ and predictions $\robot_i(z)$ are taken as fixed; the only randomness is in the assignment vector $Z$.
Such predictions typically condition on the baseline covariates; we suppress this in the notation to emphasize the generative AI use case that typically abstracts away from baseline features. These black-box predictions could come from a variety of sources, including traditional supervised learning models trained on historical data, large language models generating ``digital twins'' of experimental participants, or AI agents simulating human behavior. We explicitly allow the predictions to depend on possible treatment assignment $z$, capturing potential treatment effect heterogeneity in the black-box model. Importantly, we do not assume that these predictions are unbiased or correctly specified. We simply view them as auxiliary information that we can use to improve efficiency.

\paragraph{Extension to complex and adaptive randomization.}
While we focus on completely randomized experiments for clarity, the framework extends to stratified, clustered, and other complex randomization designs, as well as adaptive designs with always-valid confidence sequences. There is a substantial literature on regression adjustment and model-assisted estimation for such designs \citep{ding2024first,su2021model,molitor2025anytime} that can readily accommodate AI predictions.

\paragraph{Extension to super-population estimands.}
Our design-based framework targets the sample average treatment effect (SATE), with randomness coming solely from treatment assignment in the finite population. Researchers are sometimes interested in super-population estimands such as the population average treatment effect (PATE) \citep{imbens2015causal}, where units are drawn i.i.d.\ from a population and the target is $\E[Y(1) - Y(0)]$. The estimators we discuss are consistent for the PATE under standard sampling assumptions, with the usual inflation of standard errors to account for sampling variability of $X$. Recent semiparametric work on hybrid estimation, including \citet{ruan2025calm} and \citet{bartolomeis2025efficient}, develops efficient AIPW-style estimators for the PATE and related functionals that are asymptotically equivalent to calibrated regression adjustment in randomized experiments; these approaches extend more naturally to observational data and non-i.i.d.\ sampling at the cost of stronger regularity conditions.

\paragraph{Extension to textual and complex treatments.}
This framework also naturally accommodates experiments with textual or high-dimensional treatments, increasingly common in survey research and online platforms \citep{saenger2024autopersuade, lin_optimizing_2024}. When treatments are text-based---such as different message framings or policy descriptions---LLMs can interpret these directly when generating predictions, extending variance reduction to settings where traditional adjustment would require manual coding. More broadly, LLMs can serve as flexible ``codebooks'' for high-dimensional data, automatically extracting features or categorizing treatments into meaningful groups \citep{egami2023using}.

\paragraph{Notation.}
We use the following notation throughout: $\bar{Y}(z) = \frac{1}{n} \sum_{i=1}^n Y_i(z)$ denotes the finite population mean outcome under treatment $z$; $\robotbar(z) = \frac{1}{n} \sum_{i=1}^n \robot_i(z)$ is the finite population mean prediction under treatment $z$; $\bar{Y}_z^{\text{obs}} = \frac{1}{n_z} \sum_{i: Z_i = z} Y_i^{\text{obs}} $ is the observed mean outcome in treatment group $z$; and $\robotbar_z^{\text{obs}} = \frac{1}{n_z} \sum_{i: Z_i = z} \robot_i(z)$ is the sample mean prediction in treatment group $z$.

\section{Estimators}\label{sec:estimators}

\subsection{Warmup: bias-correcting AI predictions}
\label{sec:warmup-ppi}

As a warmup, we first consider estimation of one component of the ATE, the average treated potential outcome $\bar{Y}(1)$. We start with the population mean AI prediction $\robotbar(1)$ as an initial, likely biased estimate of $\bar{Y}(1)$. Following a long literature on survey calibration and control variates \citep{sarndal2003model,angelopoulos2023prediction,egami2023using,song2026demystifying}, we can augment this biased estimate with a correction term. Under random assignment, the difference between sample predictions and observed outcomes among treated units,
$$
\robotbar_1^{\text{obs}} - \bar{Y}_1^{\text{obs}},
$$
is an unbiased estimate of the prediction bias $\robotbar(1) - \bar{Y}(1)$, regardless of the prediction quality of the AI. Subtracting this estimated bias from the AI prediction yields the \emph{model-assisted} estimator \citep{sarndal2003model}:
\begin{align*}
\robothat^{\text{adj}}(1) &= \robotbar(1) - \left(\robotbar_1^{\text{obs}} - \bar{Y}_1^{\text{obs}}\right) \\ 
&= \bar{Y}_1^{\text{obs}}  + \left(\robotbar(1) - \robotbar_1^{\text{obs}}\right).
\end{align*}
This estimator, equivalent to the unweighted form of \emph{prediction-powered inference} \citep[PPI;][]{angelopoulos2023prediction} and to \emph{design-based supervised learning} \citep{egami2023using}, is unbiased for $\bar{Y}(1)$ regardless of prediction quality. If the model is perfect, the correction vanishes; if the model is poor, the correction pulls the estimate back toward the observed data.

\subsection{Model-assisted estimation for randomized experiments}
We now apply the same bias-correction strategy to each treatment arm to estimate the full average treatment effect $\tau = \bar{Y}(1) - \bar{Y}(0)$.
As before, we begin with a pure ``synthetic experiment'' estimate using only the AI predictions:
$$
\hat{\tau}_{\robot} = \robotbar(1) - \robotbar(0).
$$
In general, this estimate will be biased for $\tau$. However, we can again use the difference between sample predictions and observed outcomes within each treatment group to estimate and correct for this bias.
Bias-correcting the AI prediction separately for treated and control groups yields the \emph{model-assisted} estimator \citep{ding2024first,poulet2025ppi_for_rct,deng2023augmentation}:
\begin{align*} 
  \hat{\tau}_{\text{ma}} &=
 \left\{\robotbar(1) - \left(\robotbar_1^{\text{obs}} - \bar{Y}_1^{\text{obs}}\right)\right\} - 
  \left\{\robotbar(0) - \left(\robotbar_0^{\text{obs}} - \bar{Y}_0^{\text{obs}}\right)\right\} \\
&= \underbrace{\bar{Y}_1^{\text{obs}} - \bar{Y}_0^{\text{obs}}}_{\hat{\tau}_{\text{unadj}}} + \underbrace{\left(\robotbar(1) - \robotbar_1^{\text{obs}}\right) - \left(\robotbar(0) - \robotbar_0^{\text{obs}}\right)}_{\text{correction term}}
\end{align*}
The second line highlights that the model-assisted estimator can also be framed in terms of \emph{control variates} \citep{deng2023augmentation,ding2024first}: we augment the unbiased difference-in-means estimator with a mean-zero correction term based on the AI predictions. The correction term has mean zero by randomization, preserving unbiasedness; whether augmentation reduces variance depends on prediction quality, with the precise condition stated below.

Finally, this estimator is equivalent to the augmented inverse propensity weighted (AIPW) estimator with known (constant) propensity scores and the AI predictions $\robot_i(z)$ playing the role of the outcome model \citep{ruan2025calm,engh2025llm,bartolomeis2025efficient}. Specifically, the AIPW estimator simplifies to:
\begin{align*}
  \frac{1}{n}\sum_{i=1}^n \Bigg[\robot_i(1) - \robot_i(0) + 
  \frac{Z_i(Y_i^{\text{obs}} - \robot_i(1))}{n_1/n} - \frac{(1-Z_i)(Y_i^{\text{obs}} - \robot_i(0))}{n_0/n}\Bigg],
\end{align*}
where $n_1/n$ is the known propensity score under complete randomization. Rearranging terms recovers $\hat{\tau}_{\text{ma}}$.

\subsection{Regression adjustment for randomized experiments}

We now discuss how standard regression adjustment is a calibrated version of the model-assisted estimator. An important limitation of the model-assisted estimator is that it can \emph{increase} variance when predictions are poor \citep{freedman2008regression, ding2024first}. In particular, the model-assisted estimator reduces variance in both treatment arms if and only if
\[
   \frac{\Cov(Y(z), \robot(z))}{\Var(\robot(z))} \geq \frac{1}{2}.
\]
If predictions are standardized to have the same variance as the outcomes, this is equivalent to requiring that the correlation between outcomes and predictions within each treatment arm exceeds $1/2$. When this condition fails, which is common in empirical settings \citep{peng2025digital}, the bias correction actually \emph{inflates} variance.

Calibration resolves this variance inflation problem. Rather than fully correcting the bias, we estimate the optimal scaling via calibration coefficients
\[
  \hat{\beta}_z = \frac{\widehat{\Cov}_z(Y, \robot(z))}{\widehat{\Var}_z(\robot(z))},
\]
the within-arm OLS slope of $Y$ on $\robot(z)$ among units assigned to arm $z$, computed separately within each treatment group. The calibrated estimator is then:
\[
  \hat{\tau}_{\text{cal}} = \bar{Y}_1^{\text{obs}} - \bar{Y}_0^{\text{obs}} + \hat{\beta}_1(\robotbar(1) - \robotbar_1^{\text{obs}}) - \hat{\beta}_0(\robotbar(0) - \robotbar_0^{\text{obs}}),
\]
which we can view as a single control variate adjustment with a scaling factor based on the estimated coefficients $\hat\beta_0$ and $\hat\beta_1$ \citep{li2020rerandomization}.

A remarkable fact due to \citet{lin2013agnostic} is that we can directly estimate $\hat{\tau}_{\text{cal}}$ via the coefficient on the treatment indicator $Z$ in the following interacted regression
\[
  Y \sim 1 + Z + \dot{\robot}(Z) + Z \times \dot{\robot}(Z),
\]
where $\dot{\robot}(z) = \robot(z) - \robotbar_z^{\text{obs}}$ are group-mean-centered predictions.\footnote{As \citet{freedman2008regression} emphasized, the OLS-based estimator carries a small finite-sample bias of order $O(1/n)$. \citet{lin2013agnostic} showed that the interacted specification renders this bias asymptotically negligible while guaranteeing (weak) efficiency gains over the unadjusted estimator under arbitrary model misspecification. In what follows we treat $\hat{\tau}_{\text{cal}}$ as approximately unbiased. Readers desiring exact unbiasedness can use the closed-form bias-correction of \citet{chang2024exact}, which removes the $O(1/n)$ term while preserving the limiting distribution.}
In practice, including the interaction term often leads to negligible change, and researchers can simply fit the regression: $Y \sim 1 + Z + \robot(Z)$. This form is known as the ANCOVA adjustment for the estimated prognostic score in the causal inference literature; see, for example, \citep{hansen2008prognostic, schuler2022increasing, guo2022machinelearningvariancereduction}.
The linear regression estimator is also equivalent to a version of CUPED \citep{deng2023augmentation}, and to the calibrated version of CALM \citep{ruan2025calm}. See \citet{vdL2026calibeating} for further equivalences among calibrated estimators.

Given the regression estimator, we can construct valid confidence intervals using standard heteroskedasticity-robust standard errors \citep{lin2013agnostic,ding2024first}.
We can quantify the variance improvement via the \emph{design effect} in treatment arm $z$, the ratio of the calibrated estimator's variance to the unadjusted variance:
$$
\text{Deff}_z = \frac{\Var\left(\bar{Y}_z^{\text{obs}} + \hat{\beta}_z(\robotbar(z) - \robotbar_z^{\text{obs}})\right)}{\Var\left(\bar{Y}_z^{\text{obs}}\right)} = 1 - \rho(z,z)^2 \leq 1,
$$
where $\rho(z,z) \equiv \text{corr}(Y_i(z), \robot_i(z))$. Since $\rho(z,z)^2 \geq 0$, calibration 
never increases variance relative to the unadjusted estimator.

\paragraph{Alternative regression specifications.}
There are many potential regression specifications for adjusting for AI predictions.
Following \citet{cohen2024noharm}, we can adjust for the black-box predictions of both potential outcomes, $(\robot(0), \robot(1))$: $Y \sim 1 + Z + \dot\robot + Z \cdot \dot\robot $, where $\dot{\robot} = \left( (\robot(0) - \robotbar_0^{\text{obs}}), (\robot(1) - \robotbar_1^{\text{obs}}) \right)$. This has a design effect of $\text{Deff}_z = 1 - \rho_{\text{mr}}(z)^2$, where  $\rho_{\text{mr}}(z)$  denotes the multiple correlation between $Y(z)$ and both predictions $(\robot(0), \robot(1))$.\footnote{Alternatively, we can adjust for only a single black-box prediction, either $\robot(0)$ or $\robot(1)$. Using $\robot(z')$ for treatment arm $z$ has a design effect of $\text{Deff}_z = 1 - \rho(z,z')^2$.}
Many other variants are common in the literature on heterogeneous treatment effect estimation; see \citet{wager2024causal} for a review.

\paragraph{Digital twin estimator.} We briefly discuss an alternative variance reduction approach based on directly predicting the missing potential outcome with ``digital twins'' \citep{argyle2023out, horton2023large, schuler2022increasing,alam2024digitaltwin}. 
This is known as the \emph{imputation} or matching estimator \citep{ding2024first,lin2023nn_ecma}.
First, construct unit-level treatment effect estimates by imputing the missing potential outcome using the AI prediction:
\[
  \hat{\tau}_i = \begin{cases}
    Y_i^{\text{obs}} - \robot_i(0) & \text{if } Z_i = 1 \\
    \robot_i(1) - Y_i^{\text{obs}} & \text{if } Z_i = 0.
  \end{cases}
\]
Averaging over all units yields the \emph{digital twin estimator}:
\[
  \hat{\tau}_{\text{dt}} = \frac{1}{n}\sum_{i=1}^n \hat{\tau}_i = \frac{n_1}{n}\left(\bar{Y}_1^{\text{obs}} - \robotbar_1(0)\right) - \frac{n_0}{n}\left(\bar{Y}_0^{\text{obs}} - \robotbar_0(1)\right),
\]
where $\robotbar_z(z') = \frac{1}{n_z} \sum_{i: Z_i = z} \robot_i(z')$ is the sample mean prediction for units in group $z$ under treatment $z'$.

This estimator is biased unless the predictions are correct on average: $\robotbar(z) = \bar{Y}(z)$ for $z \in \{0,1\}$. This is a strong assumption that LLMs and other AI models rarely satisfy; indeed, if it were true, there would be no need for non-synthetic data. This makes the digital twin estimator risky in practice despite its intuitive appeal.

\paragraph{Nonlinear calibration.}
For binary or count outcomes, continuous AI predictions $(\robot(0), \robot(1))$ can be calibrated nonlinearly, for example, by including them in a logistic or Poisson regression, or fitting a generalized additive model. The generalized Oaxaca-Blinder framework \citep{basse2024oaxaca}, with an extension from \citet{cohen2024noharm}, shows that such nonlinear calibration inherits the same robustness properties: the resulting estimator is unbiased regardless of calibration model quality and reverts to the unadjusted estimator when predictions are uninformative.

\subsection{Comparison of estimators}

Table~\ref{tab:estimators-deff} summarizes the bias and design effect for each estimator discussed above.

\begin{table*}[t]
\centering
\small
\renewcommand{\arraystretch}{1.5}
\begin{tabular}{@{}l c l c@{}}
\toprule
Estimator & Unbiased? & Design effect, $\mathrm{Deff}_z$ & Always $\leq 1$? \\
\midrule
Unadjusted & Yes & $1$ & --- \\
Digital twin & No & $4p(1-p)\frac{n_z}{n}(1 - \rho(z, z))$ & Only if balanced \\
Model-assisted & Yes & $2(1- \rho(z,z))$ & Only if $\rho > 0.5$ \\
Regression with $\robot(Z)$ & Approx. & $1-\rho(z,z)^2$ & Yes \\
Regression with $\{\robot(0), \robot(1)\}$  & Approx. & $1-\rho_{\mathrm{mr}}(z)^2$ & Yes \\
\bottomrule
\end{tabular}
\caption{Bias and design effect for each estimator. Design effects assume predictions are standardized to match outcome variance.
Regression adjustment has a very small finite-sample bias that vanishes in most realistic settings \citep{lin2013agnostic}.}
\label{tab:estimators-deff}
\end{table*}

\paragraph{Calibration guarantees robustness.} The calibrated estimators always achieve $\text{Deff}_z \leq 1$; that is, incorporating AI predictions does not \emph{increase} variance. When predictions are uninformative ($\rho = 0$), the design effect equals 1 and we simply recover the unadjusted estimator's precision. When predictions are highly correlated with outcomes, variance drops substantially. 

\paragraph{Uncalibrated approaches are fragile.} 
Uncalibrated estimators only achieve efficiency gains with very strong predictions. For instance, the model-assisted estimator improves over the baseline within arm $z$ only when $\rho(z,z) > 0.5$, and otherwise \emph{inflates} that arm's variance. The digital twin estimator can reduce variance with weaker correlations, but introduces bias that may be arbitrarily large. 

\paragraph{Both predictions maximize gains.} When the model generates predictions for both potential outcomes $(\robot(0), \robot(1))$, adjusting for both will typically achieve greater variance reduction than adjusting for the observed prediction alone, $\robot(Z_i)$. The additional gain beyond the factual prediction depends on whether the counterfactual prediction $\robot(1-z)$ provides information about $Y(z)$ not already captured by $\robot(z)$.

\section{Implementation details}\label{sec:implementation}

We briefly discuss how to generate the AI predictions, $\robot_i(z)$.

\paragraph{Black-box models.} We consider three prediction approaches:
\begin{itemize}
  \item \emph{Large language models (LLMs):} Commercially available models such as GPT-5.4 that generate outcome predictions zero-shot from unit descriptions (e.g., persona profiles, baseline characteristics) and treatment conditions. LLMs are particularly well-suited when treatments are textual or when rich unstructured context is available. 
  \item \emph{Tabular foundation models:} TabPFN \citep{hollmann2022tabpfn}, a transformer pretrained on synthetic tabular datasets, which conditions on the labeled dataset at inference time (few-shot by construction) and requires no hyperparameter tuning.
  \item \emph{Standard supervised learning:} XGBoost and Ridge regression, trained on engineered features with cross-validated hyperparameters; we use Ridge rather than Lasso to provide stable shrinkage at the small per-fold sample sizes encountered under cross-fitting.
\end{itemize}

\paragraph{Generating predictions.}
To maximize variance reduction, we query the model separately under each treatment condition to obtain both $\robot_i(0)$ and $\robot_i(1)$. Prompt design can substantially affect prediction quality; pilot data can guide prompt engineering by optimizing for the within-arm correlation $\rho(z,z)$, though analysts should be cautious about overfitting the prompt to a small pilot sample.

\paragraph{Cross-fitting.} Zero-shot LLM predictions are constructed without access to the realized $Y^{\text{obs}}$ or $Z^{\text{obs}}$. We can therefore view $\robot_i$ as if it were a fixed pre-treatment covariate, no different from age or baseline survey responses. Thus standard regression adjustment applies with no cross-fitting or sample-splitting required. By contrast, models trained on experimental outcome data (TabPFN, XGBoost, Ridge, fine-tuned LLMs) require cross-fitting to preserve design-based validity \citep{guo2022machinelearningvariancereduction}; in our experiments, we use 5-fold cross-fitting for all ML baselines. Nonlinear calibration models fit on top of predictions may also require cross-fitting when the calibration function is estimated from the data.

\paragraph{Combining prediction sources.} 
LLMs can also \emph{featurize} unstructured data, extracting embeddings or categorizing free-text responses, ${\bm{X}}_\robot$ \citep{engh2025llm, egami2023using}. These can then be fed into downstream models such as TabPFN or used alongside the AI predictions as additional baseline covariates:
\[
  Y \sim 1 + Z + \dot{\bm{X}}_\robot + Z \cdot \dot{\bm{X}}_\robot + \dot{\robot}(Z) + Z \cdot \dot{\robot}(Z),
\]
where $\dot{\bm{X}}_\robot$ denotes group-mean-centered AI features. The resulting estimator is approximately unbiased and achieves a design effect of $1 - \rho_{\text{mr}}(z)^2$, where $\rho_{\text{mr}}(z)$ is now the multiple correlation between $Y(z)$ and the full covariate and prediction vector. Including both can only improve (or maintain) precision relative to using either alone, except when features are very high-dimensional \citep[see][]{lei2021regression}.

\paragraph{Binary and discrete outcomes.}
For binary outcomes, continuous AI predictions can yield greater variance reduction than binarized predictions (Section~\ref{sec:simulations}). Directly eliciting a probability $\robot \in [0,1]$ from the LLM is one option, though such probabilities can be poorly calibrated. We therefore favor two alternatives that yield a well-scaled continuous score: extracting the \emph{log-probability} of the positive-class token (e.g., ``Yes''), which needs no extra queries, or \emph{averaging multiple completions} sampled at positive temperature ($10$--$20$ draws, batchable in a single call). See Appendix~\ref{app:prediction-methods} for comparisons.

\section{Controlled synthetic experiments}\label{sec:simulations}
We run a series of controlled synthetic experiments to assess the plausible range of precision gains from AI assistance and how this varies with both the quality of the AI and the difficulty of the prediction task.

In each set of experiments, we generate a finite population of $n = 1,000$ units with covariates $X_i \sim \text{Uniform}(0, 1)$ for $i = 1, \ldots, n$. We consider a completely randomized design with $n_1 = n_0 = 500$ units assigned to treatment and control, respectively.
In the first set of experiments, we generate potential outcomes under control as $Y_i(0) = m_0(X_i) + \varepsilon_i$ where $\varepsilon_i \overset{iid}{\sim} \mathcal{N}(0, \sigma^2_y)$ and $m_0(x) =  \sin(3 \cdot \pi \cdot x) / (3 \cdot \pi \cdot x)$ is a non-linear but smooth function. We then generate potential outcomes under treatment as $Y_i(1) = Y_i(0) + \tau + X_i$, where $\tau$ is chosen so that the average treatment effect is 0. Note that the potential outcomes are perfectly correlated conditional on the covariate $X$, due to the shared noise term $\varepsilon_i$. The variance of this noise term, $\sigma^2_y$, controls the inherent difficulty of the prediction task: lower values correspond to easier prediction problems. We use $R^2_m$ to denote the proportion of variance in $Y(0)$ explained by the true underlying model $m_0(X)$, i.e., $R^2_m = \Var(m_0(X)) / (\Var(m_0(X)) + \sigma^2_y)$. Higher values of $R^2_m$ correspond to tasks where the outcome is more predictable from the covariate.

To simulate AI predictions, we generate predicted potential outcomes as $\robot_i(0) = m_0(X_i) + b + \eta_i$ and $\robot_i(1) = \robot_i(0) + \tau + X_i$, where $\eta_i \overset{iid}{\sim} \mathcal{N}(0, \sigma^2_{\robot})$ is independent noise and $b$ is a bias term that reflects systematic over- or under-prediction. We use $R^2_{\robot}$ to denote the signal-to-total-variance ratio of the AI prediction, $R^2_{\robot} = \Var(m_0(X)) / (\Var(m_0(X)) + \sigma^2_{\robot})$; higher values correspond to more accurate AI predictions. The squared correlation between $Y(0)$ and $\robot(0)$, which directly governs variance reduction, is $\text{corr}(Y(0), \robot(0))^2 = R^2_m \cdot R^2_{\robot}$. Because the AI predictions share the same functional form as the true outcome model, idiosyncratic errors in the AI predictions come from the noise term $\eta_i$, and so $\sigma^2_{\robot}$ controls the quality of the AI predictions.

For the second and third sets of experiments, we consider the case where outcomes are binarized so that $\tilde{Y}_i(z) = \mathbbm{1}\{Y_i(z) \geq 0\}$. In the second set of experiments, the AI predictions are also binarized, so that $\tilde{\robot}_i(z) = \mathbbm{1}\{\robot_i(z) \geq 0\}$. In the third set of experiments, the AI predictions remain continuous.

\begin{figure}[htbp]
  \centering
  \begin{subfigure}[t]{0.45\textwidth}
    \centering
    \includegraphics[width=\linewidth]{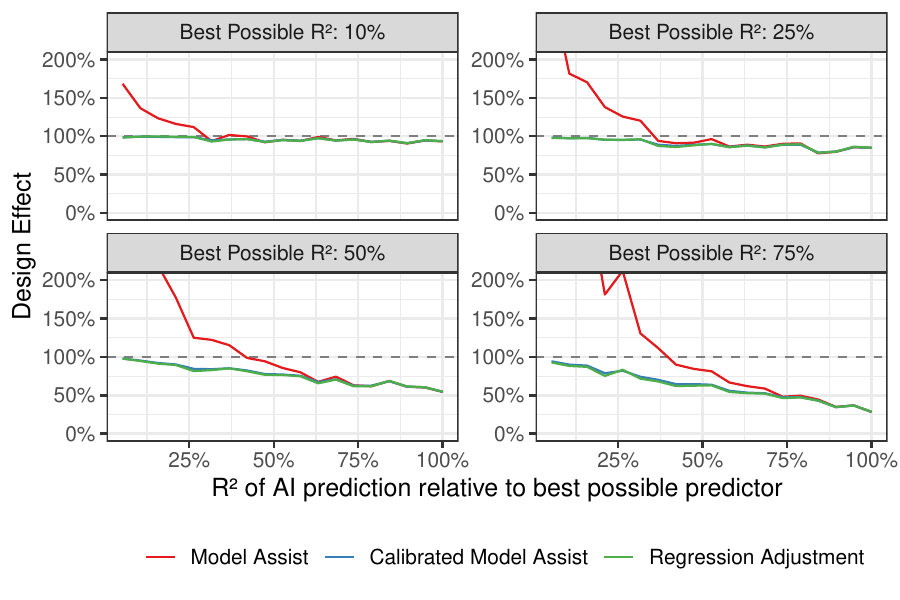}
        \caption{Continuous outcomes}
  \end{subfigure}\\[3em]
  \begin{subfigure}[t]{0.45\textwidth}
    \centering
    \includegraphics[width=\linewidth]{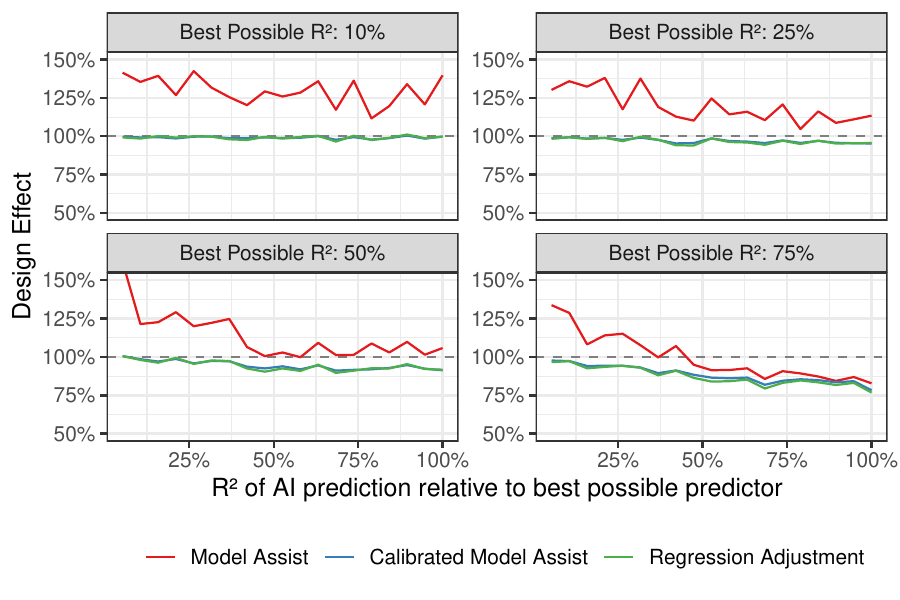}
        \caption{Binary outcomes and AI predictions}
  \end{subfigure}\\[3em]
  \qquad \begin{subfigure}[t]{0.45\textwidth}
    \centering
    \includegraphics[width=\linewidth]{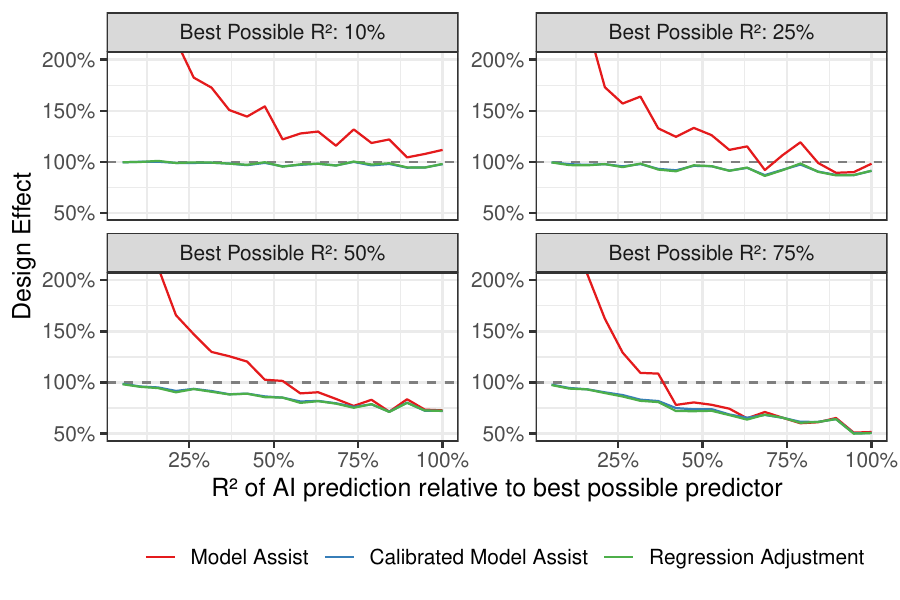}
        \caption{Binary outcomes and continuous AI predictions}
  \end{subfigure}
  \caption{Design effect from AI-assisted variance reduction across a range of prediction quality ($R^2_{\robot}$) and outcome predictability ($R^2_m$). Results are shown for continuous outcomes (a), binary outcomes with binarized AI predictions (b), and binary outcomes with continuous AI predictions (c).}
  \label{fig:sim_efficiency}
\end{figure}

Figure~\ref{fig:sim_efficiency} shows the design effect, $\frac{\Var(\hat{\tau}_{\text{adj}})}{\Var(\hat{\tau}_{\text{unadj}})}$, from AI-assisted variance reduction across a range of adjustment techniques, prediction quality ($R^2_{\robot}$) and outcome predictability ($R^2_m$).
(Here and below we report the overall design effect rather than separate per-arm design effects.)
For continuous outcomes (a), we see that when the underlying outcomes are difficult to predict (low $R^2_m$), even AI predictions that are as predictive as possible (i.e., $R^2_{\robot}$ close to $R^2_m$) yield only modest efficiency gains, while larger gains are possible when both the outcome is very predictable (high $R^2_m$) and the AI predictions are of high quality ($R^2_{\robot}$ close to $R^2_m$). Consistent with the theoretical results in Table~\ref{tab:estimators-deff}, the uncalibrated model-assisted estimator inflates variance at low $R^2_{\robot}$, while linear regression adjustment maintains $\text{Deff} \leq 1$ throughout.

For case (b), with both binary outcomes and AI predictions, we see that the potential efficiency gains are even more muted. This is because even if the underlying continuous outcomes are highly predictable, the binary outcome may still have high entropy when the continuous outcome is near the binarization threshold. Binary AI predictions exacerbate this issue. However, when the AI predictions remain continuous (c), we see that efficiency gains are more similar to the continuous outcome case, although attenuated somewhat due to higher variability of the outcomes.

Appendix~\ref{app:sim-bias} reports complementary simulations varying the AI-prediction bias $b$. Varying $b$ symmetrically across arms, the uncalibrated model-assisted estimator's variance inflates rapidly with $|b|$, while calibrated regression adjustment is unaffected (Figure~\ref{fig:varying-bias}). Appendix~\ref{app:sim-aipw} further compares against an AIPW estimator with estimated propensity score under both symmetric and \emph{asymmetric} bias (where only the estimated control outcome is shifted): calibrated regression is the only estimator that remains simultaneously unbiased, well-calibrated, and powerful across regimes. By contrast, AIPW is point-unbiased but substantially overcovers;  the digital twin fails under asymmetric bias.

\section{Empirical validation}
\subsection{Digital Twins}\label{sec:empirical-twins}

Our first empirical comparison uses Twin-2K-500 Mega-Study~\citep{peng2025digital}, which constructed digital twins of 2,000 participants by having GPT-4 simulate responses based on detailed persona profiles derived from 500 prior survey questions.

We analyze two randomized experiments that test opt-in vs.\ opt-out framing, \emph{Organ Donation} and \emph{Green Energy} ($n{=}600$ per experiment, approximately balanced across arms). For each participant we observe the actual outcome under the realized treatment and generate LLM predictions under both treatment conditions.

We generate predictions using GPT-5.4 following the Twin-2K-500 protocol, providing each participant's full persona profile (${\sim}$127K characters) without truncation. We extract continuous predictions from token logprobs in a single API call per persona-condition pair, obtaining $\robot_i(z) \in [0,1]$ as the model's estimated probability of the target behavior (Appendix~\ref{app:prediction-methods}).

For ML baselines, we engineer 26 composite psychological scale scores from the raw survey items, following \citet{peng2025digital}, and use GPT-5.4 to select 10 theoretically relevant features per outcome. Importantly, we do so \emph{without using any outcome data}, avoiding data leakage. All ML models (Ridge, XGBoost, TabPFN~\citep{hollmann2022tabpfn}) use T-Learner with leave-one-out cross-validation. Appendix~\ref{app:twin-details} gives full details on feature engineering, model specifications, and cross-fitting.

\begin{table}[t]
\centering
\small
\caption{Estimator comparison on Twin-2K-500 experiments ($n{=}600$). VR\% = variance reduction relative to unadjusted. Negative values indicate variance \emph{increase}. The Digital twin estimator produces biased estimates. The Model-assisted estimator increases variance when $\rho < 0.5$, consistent with our theoretical analysis.}
\label{tab:twin-empirical}
\resizebox{\linewidth}{!}{%
\setlength{\tabcolsep}{3pt}
\begin{tabular}{l cc cc}
\toprule
 & \multicolumn{2}{c}{Organ Donation} & \multicolumn{2}{c}{Green Energy} \\
\cmidrule(lr){2-3} \cmidrule(lr){4-5}
Estimator & $\hat{\tau}$ (SE) & VR\% & $\hat{\tau}$ (SE) & VR\% \\
\midrule
Unadjusted & 0.044 (0.040) & --- & 0.247 (0.040) & --- \\
Digital twin & $-$0.278 (0.029) & $+$49.1\% & $-$0.196 (0.037) & $+$12.7\% \\
Model-assisted & 0.072 (0.047) & $-$36.3\% & 0.269 (0.041) & $-$8.0\% \\
Reg. with $\robot(Z)$ & 0.044 (0.040) & $+$2.1\% & 0.247 (0.037) & $+$13.7\% \\
Reg. with $\{\robot(0), \robot(1)\}$ & 0.044 (0.040) & $+$2.7\% & 0.247 (0.037) & $+$14.3\% \\
Reg. with $\{\robot(0), \robot(1), X\}$ & 0.044 (0.039) & $+$4.5\% & 0.247 (0.036) & $+$16.4\% \\
\bottomrule
\end{tabular}%
}

\end{table}

Table~\ref{tab:twin-empirical} presents the main results, which align with our discussion above.
The unbiased difference-in-means estimates are $\hat{\tau} \approx 0.04$ and $\hat{\tau} \approx 0.25$ for Organ Donation and Green Energy, respectively.
The \emph{digital twin} estimators show substantial bias, with $\hat{\tau} = -0.28$ for Organ Donation and $\hat{\tau} = -0.20$ for Green Energy.
The uncalibrated \emph{model-assisted} estimator \textit{increases} variance by 36\% for Organ Donation ($\rho \approx 0.13$) and 8\% for Green Energy ($\rho \approx 0.37$), consistent with our result that this estimator requires $\rho > 0.5$ to reduce variance. By contrast, the standard regression estimator achieves 13.7\% variance reduction for Green Energy and 2.1\% for Organ Donation. Conditioning on both $\robot(0)$ and $\robot(1)$ slightly improves variance for both, achieving 14.3\% vs.\ 13.7\% for Green Energy and 2.7\% vs.\ 2.1\% for Organ Donation. Additionally adjusting for baseline covariates (27 composite scales from the pre-treatment survey wave) further improves variance, with a 16.4\% reduction for Green Energy and 4.5\% for Organ Donation.


\begin{table}[t]
\centering
\small
\caption{LLM vs.\ ML with LLM-extracted features. $\rho$ = average within-arm correlation. VR = variance reduction (\%).}
\label{tab:ml-vs-llm}

\begin{tabular}{lrrrr}
\toprule
 & \multicolumn{2}{c}{Organ Donation} & \multicolumn{2}{c}{Green Energy} \\
\cmidrule(lr){2-3} \cmidrule(lr){4-5}
Predictor & $\rho$ & VR\% & $\rho$ & VR\% \\
\midrule
LLM with binary pred.   & .11 & +2\% & {.36} & {+13\%} \\
LLM with logprobs & .13 & +2\% & {.37} & {+14\%} \\
Ridge with LLM-selected features     & .20 & +4\% & .38 & +14\% \\
Ridge with logprobs \& features      & .20 & +4\% & .40 & +16\% \\
\bottomrule
\end{tabular}

\end{table}

Table~\ref{tab:ml-vs-llm} compares LLM predictions against a ridge regression with 10 LLM-selected features.
For the Organ Donation study, LLM predictions are between $\rho = 0.11$ and $\rho = 0.13$, depending on whether predictions are based on the extracted log-probabilities or on averaging over many binary predictions. Ridge regression with the LLM-extracted features outperforms these, with $\rho = 0.20$. For the Green Energy study, where the GREEN attitude scales directly measure the construct of interest, all methods perform comparably, with $\rho$ between 0.36 and 0.38.
Combining the LLM logprob predictions with the LLM-selected features in a single ridge regression yields a small further gain on Green Energy ($\rho = 0.40$, +16\% VR) and no detectable gain on Organ Donation ($\rho = 0.20$, +4\% VR), confirming the substantial overlap between the two signals.
Thus, using LLMs as a pre-processing step to extract relevant features can be a competitive alternative to using the AI predictions directly.

\subsection{Hillstrom Email Marketing Analysis}\label{sec:hillstrom}

We next evaluate variance reduction in a traditional marketing setting with tabular covariates, using the Hillstrom MineThatData E-mail Analytics dataset. The dataset contains 64,000 customers randomly assigned to one of three treatments (1/3 each), referred to as: \emph{Men's email}, \emph{Women's email}, or Control (no email). We analyze \emph{Men's Email} vs.\ Control ($n \approx 42{,}613$) for the Visit outcome. Each customer has 8 pre-treatment covariates capturing purchase history, demographics, and channel preferences (Appendix~\ref{app:hillstrom-details}).
For LLM predictions, we convert tabular features to structured text and use GPT-5.4 with direct probability elicitation, prompting for a numeric 0--100 probability estimate rather than binary classification. We found this produces better-calibrated predictions than binary or logprobs modes, which were overconfident and yielded near-zero variance reduction (see Appendix~\ref{app:hillstrom-details} for prompt design).

\begin{table}[t]
\centering
\small
\caption{Hillstrom Visit Prediction, Mens Email vs.\ Control ($n{=}42{,}613$). ML methods outperform LLM when tabular covariates directly capture the relevant signal.}
\label{tab:hillstrom-comparison}
\begin{tabular}{lrr}
\toprule
Method & Avg $\rho$ & VR (\%) \\
\midrule
Ridge with $X$ & 0.170 & 2.9 \\
XGBoost with $X$ & 0.148 & 2.2 \\
LLM direct prob. & 0.105 & 1.1 \\
\bottomrule
\end{tabular}
\end{table}

Table~\ref{tab:hillstrom-comparison} presents the results. 
Ridge and XGBoost on the original features both yield reasonable predictions: with $\rho = 0.170$ (2.9\% VR) for ridge and $\rho = 0.148$ (2.2\% VR) for XGBoost.
By contrast, the direct probability predictions from an LLM (GPT-5.4) achieve $\rho = 0.105$ (1.1\% VR).
Unlike with the previous Green Energy and Organ Donation studies, here eight tabular features directly measure purchase behavior, the primary driver of future visits, and ML methods outperform the LLM by more than a factor of two.


\subsection{Large-Scale Online Platform Search Experiment}\label{sec:tech-experiment}

Finally, we evaluate a large-scale randomized experiment conducted on a consumer-facing online platform. 
The experiment tested a new search ranking algorithm against an incumbent ranker, with users randomly assigned to see search results generated by one of the two.
In this setting, key business outcomes, such as successful downstream conversions, are sparse and highly variable at the user level, and treatment effects are typically small. 
Achieving adequate statistical power often requires large sample sizes and long experiment runtimes, making this a natural testbed for variance reduction techniques.

\paragraph{Data and experimental design.}
Users who searched on the platform during the experiment were randomly assigned at the user level, with equal probability, to treatment or control. 
Treatment users were shown search results based on the new ranking algorithm, while control users continued to see search results from the incumbent ranker.
Assignment was persistent for the duration of the experiment, so each user consistently experienced the same ranker across all search sessions.

The experiment ran for one month and included 13.2 million users, who entered on a rolling basis. 
From this population, we construct a focused analysis cohort by randomly sampling $n_{1} = 2500$ treated users and $n_{0} = 2500$ control users who entered the experiment on its first day. 
We analyze this fixed subsample to facilitate a direct comparison of estimator performance under limited sample sizes. 
Restricting to users with a common entry time also simplifies the estimand by ensuring that all users are exposed to the assigned treatment for the full experiment duration, 
rather than a mixture of exposure lengths.

For each user, we observe a pre-treatment outcome $Y_{i, \text{pre}}$, a rich set of pre-treatment user activity $\bm{X}_i$, and an LLM-based prediction of conversion likelihood $\robot_i(0) \in [0,1]$ constructed from pre-treatment user activity.

The outcome of interest is the number of successful downstream conversions per user.
This metric is highly zero-inflated, with the large majority ($93\%$) of users having no more than one conversion during the experiment; most users ($70\%$) have zero conversions.
We include implementation and evaluation details in Appendix~\ref{app:tech-experiment-details}, including the construction of LLM predictions of user conversion from raw activity data.

\begin{table}[t]
\centering
\small
\setlength{\tabcolsep}{4pt}
\renewcommand{\arraystretch}{1.2}
\caption{Impact of a new search ranking algorithm on successful downstream conversions ($n_1 = 2500$, $n_0 = 2500$). We report the estimated average treatment effect ($\hat{\tau}$), standard error, and $p$-value for several estimators. Deff is the empirical design effect (ratio of adjusted to unadjusted variance), and VR is variance reduction ($\%$). LLM predictions of downstream conversion $\robot(0)$ constructed from pre-treatment information yield only small efficiency gains on their own and add little beyond the pre-treatment outcome $Y_{pre}$.}
\label{tab:tech-results}
\begin{tabular}{l c c c c}
\toprule
Estimator & $\hat{\tau}$ (SE) &  pval & Deff & VR(\%)\\
\midrule
Unadjusted & 0.0380 (0.0250)&  0.13&1  & --- \\
Reg. with $\robot(0)$ & 0.0380 (0.0249)&  0.13&0.99 &1.3\%\\
Reg. with $Y_{\text{pre}}$ & 0.0380 (0.0230)&  0.10&0.85&15.5\%\\
Reg. with $Y_{\text{pre}}$, $\robot(0)$ & 0.0380 (0.0229)& 0.10& 0.84&16.4\%\\
\bottomrule
\end{tabular}
\end{table}

\paragraph{Results.}
Table~\ref{tab:tech-results} shows the results for the unadjusted estimator and various regression estimators. (Due to data limitations, we are unable to report other estimators for this experiment.)
The Unadjusted estimator finds a positive but non-significant effect of the new search ranking algorithm on downstream conversions ($\hat{\tau} = 0.0380$, SE $= 0.0250$, pval $= 0.13$). 
Regression with $Y_{\text{pre}}$ is equivalent to the standard CUPED estimator~\citep{deng2013improving, deng2023augmentation} used in large-scale industry A/B testing platforms, where the pre-treatment value of the outcome metric (e.g., a user's prior conversion count) is used for covariate adjustment. This industry-standard estimator provides modest efficiency gains, with a $15.5\%$ reduction in variance.
By contrast, regression with the pre-treatment LLM prediction alone, $\robot(0)$, yields only a $1.3\%$ variance reduction, and adding it on top of $Y_{\text{pre}}$ contributes little beyond CUPED ($16.4\%$). 
In this setting, the pre-treatment LLM prediction provides limited additional signal over the pre-treatment outcome itself, and the estimated effect remains inconclusive after adjustment.

\section{Discussion}
\label{sec:discussion}

We argue that linear regression adjustment is a simple and effective way to incorporate AI predictions into randomized experiments, building on a long causal inference literature on covariate and prognostic score adjustment. We can also motivate regression adjustment as a calibrated version of recently popular bias-corrected estimators, such as prediction-powered inference. Regression has an appealing ``do no harm'' property: it remains unbiased regardless of prediction quality and (weakly) reduces variance relative to the unadjusted estimator. By contrast, uncalibrated approaches such as model-assisted and digital twin estimators can inflate variance when predictions are poor, and digital twin estimators can also be arbitrarily biased.
In applications with substantial text and other unstructured data, LLM predictions can outperform standard tabular models; in other settings, the gains are often modest and context-dependent. In our technology platform application, for example, LLM predictions based on pre-treatment user activity added little beyond the pre-treatment outcome itself. Regardless of prediction quality, however, researchers can always incorporate this information into existing covariate adjustment frameworks alongside other baseline covariates.

While regression calibration protects against variance inflation and model misspecification, it does not by itself address other important guardrails around AI deployment. The most prominent is to avoid pipeline failures from data leakage, which invalidate the statistical guarantees of the design-based framework and can lead to biased estimates. In particular, researchers should ensure that AI predictions are generated from pre-treatment information only and have no access to the realized treatment assignments or outcomes. A related danger is the use of AI tools to enable $p$-hacking by repeatedly tuning prompts or other LLM choices; here too, researchers should pre-specify any adjustment procedure to maintain validity.

We suggest several directions for future work:

\paragraph{Subgroup analysis with predicted treatment effects.}
AI predictions under both treatment conditions yield unit-level predicted treatment effects $\robot_i(1) - \robot_i(0)$, which can define subgroups for heterogeneous effect analysis. LLMs can also generate otherwise latent covariates, such as imputed party affiliation, user intent, or personality traits, that serve as subgroup moderators, analogous to using LLMs for survey imputation \citep{egami2023using}. While promising, a key danger is that such LLM-generated moderators might be biased, which would complicate estimates of treatment effect heterogeneity. 

\paragraph{Surrogate outcomes.}
When treatment effects on the primary outcome are slow to materialize, AI predictions may serve as surrogate outcomes that enable earlier decisions. Conversion likelihood scores, such as those in our technology platform experiment (Section~\ref{sec:tech-experiment}), are a candidate when they can be computed early and sequentially, and reliably track downstream behavior. A promising direction is to formalize this connection to the surrogacy literature and to characterize when such scores add signal beyond pre-treatment outcomes.

\paragraph{Randomization-based testing with statistics based on AI predictions.}
Our estimators target point estimation and Wald-type intervals, but the same design-based logic extends to Fisher-style randomization tests. For instance, \citet{guo2025ml} construct ML-assisted randomization tests that compare cross-validation errors of models with and without the treatment indicator, retaining finite-sample validity under the sharp null while gaining power from flexible learners. A natural complement to the regression adjustment approach we consider is to use AI predictions to construct similarly informed test statistics or to define more powerful covariate-adjusted randomization tests.

\paragraph{Mixed subject designs.}
A recent line of work, including the \emph{mixed subjects} design of \citet{broska2025mixed}, instead treats AI predictions as ``additional subjects'' alongside human participants. The implied estimand then averages over a mixed population of human and synthetic units rather than the human-only SATE. A natural way forward is to treat this as a problem of generalizability: estimate the human-only effect as discussed here, then formally transport it to the mixed target population, drawing on the literature on transporting treatment effects across populations \citep{degtiar2023review}. This framing keeps the target estimand and its assumptions explicit.

\bibliographystyle{plainnat}
\bibliography{references}
\appendix
\section{Derivations of Estimator Properties}\label{app:derivations}
This appendix provides additional details on the bias and variance results summarized in Table~\ref{tab:estimators-deff}. Throughout, we define $\kappa(z, z') = \Var(\robot(z'))/\Var(Y(z))$ as the ratio of prediction variance to outcome variance.

\paragraph{Digital twin estimator.}
The digital twin estimator is biased unless the model is correct on average:
$
  \E[\hat{\tau}_{\text{dt}}] - \tau = \frac{n_0}{n} \left( \robotbar(1) - \bar{Y}(1) \right) - \frac{n_1}{n} \left(\robotbar(0) -  \bar{Y}(0) \right).$
The variance is:
\begin{align*}
  \Var(\hat{\tau}_{\text{dt}}) & = \Var\left(\frac{1}{n}\sum_{i=1}^n Z_i \left(Y_i(0) + Y_i(1) - (\robot_i(1) - \robot_i(0))\right)\right)\\
  & \leq \frac{2p(1-p)}{n} \left(\Var(Y(0) - \robot(0)) + \Var(Y(1) - \robot(1))\right).
\end{align*}

The design effect for each treatment group is:
\begin{align*}
  \text{Deff}_z &= \frac{\frac{2p(1-p)}{n} \Var(Y(z) - \robot(z))}{\Var(\bar{Y}_z^{\text{obs}})}\\
  &= 2p(1-p)\frac{n_z}{n}\left(1 + \sqrt{\kappa(z, z)}\left(\sqrt{\kappa(z, z)} - 2 \rho(z, z)\right)\right).
\end{align*}

If predictions are standardized to have the same variance as outcomes ($\kappa(z, z) = 1$):
$
  \text{Deff}_z(\hat{\tau}_{\text{dt}}) = 4p(1-p)\frac{n_z}{n}(1 - \rho(z, z)).
$

\paragraph{Model-assisted estimator.}
The model-assisted estimator is unbiased for any model. Its variance is:
\begin{align*}
  \Var\left(\hat{\tau}_{\text{ma}}\right) & = \Var\left(\bar{Y}_1^{\text{obs}} - \robotbar_1^{\text{obs}} - (\bar{Y}_0^{\text{obs}} - \robotbar_0^{\text{obs}})\right)\\
  & \leq \frac{1}{n_1} \Var(Y(1) - \robot(1)) + \frac{1}{n_0} \Var(Y(0) - \robot(0)).
\end{align*}

The design effect in each treatment group is:
\begin{align*}
  \text{Deff}_z &= \frac{\Var\left(Y(z) - \robot(z)\right)}{\Var(Y(z))}
  = 1 + \sqrt{\kappa(z, z)}\left(\sqrt{\kappa(z, z)} - 2 \rho(z, z)\right).
\end{align*}

With standardized predictions ($\kappa(z, z) = 1$):
$
  \text{Deff}_z(\hat{\tau}_{\text{ma}}) = 2(1 - \rho(z, z)).
$
This is only less than 1 if $\rho(z, z) > 0.5$.

\paragraph{Calibrated model-assisted estimator.}
The calibrated estimator is unbiased for any model. With oracle calibration coefficient $\beta_z^\ast = \Cov(Y(z), \robot(z))/\Var(\robot(z))$, the design effect is:
\begin{align*}
  \text{Deff}_z &= \frac{\Var\left(Y(z) - \beta_z^\ast \robot(z)\right)}{\Var(Y(z))}\\
  &= 1 + (\beta_z^\ast)^2 \frac{\Var(\robot(z))}{\Var(Y(z))} - 2 \beta_z^\ast \frac{\Cov(Y(z), \robot(z))}{\Var(Y(z))} = 1 - \rho(z, z)^2.
\end{align*}

The feasible estimator using $\hat{\beta}_z$ is asymptotically equivalent to the oracle version since:
\[
  \hat{\tau}_{\text{cal}} - {\tau}_{\text{ocal}}^* = (\hat{\beta}_1 - \beta_1^\ast)(\robotbar(1) - \robotbar_1^{\text{obs}}) - (\hat{\beta}_0 - \beta_0^\ast)(\robotbar(0) - \robotbar_0^{\text{obs}}) = O_p(1/n).
\]






\section{Additional simulations}\label{app:sim-bias}

\begin{figure}[htbp]
  \centering
  \includegraphics[width=0.35\textwidth]{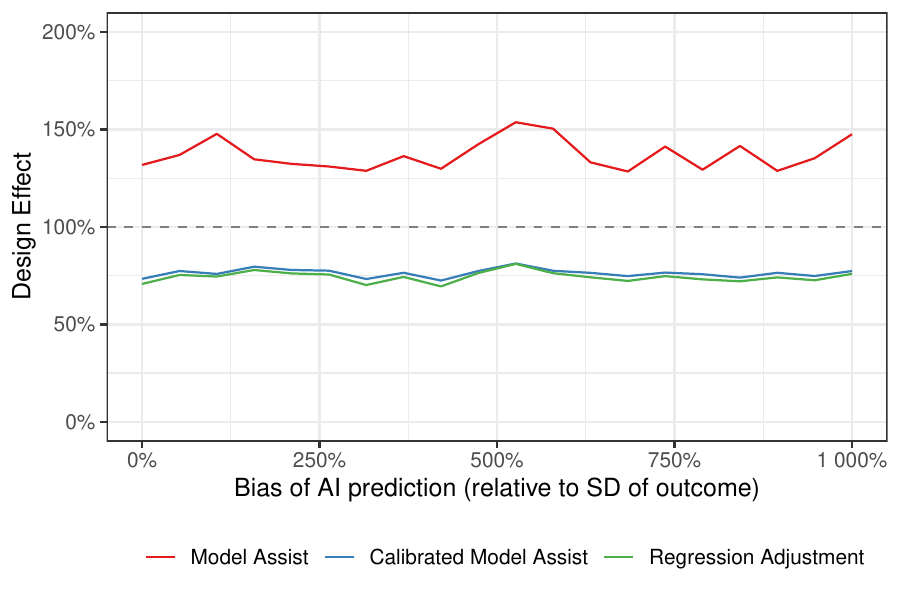}
  \caption{Design effect as a function of the AI prediction bias $b$, relative to the outcome standard deviation.}
  \label{fig:varying-bias}
\end{figure}

\paragraph{Varying prediction bias.}
The simulations in Section~\ref{sec:simulations} fix the AI-prediction bias parameter $b = 0$. Here we vary $b$ from 0 to 1 (in units of the outcome standard deviation) to illustrate how each estimator responds to systematic over- or under-prediction by the AI. Figure~\ref{fig:varying-bias} reports the design effect for the Model Assist, Calibrated Model Assist, and (calibrated) Regression Adjustment estimators at fixed $R^2_m$ and $R^2_\robot$.

Consistent with the theory in Section~\ref{sec:estimators}, the uncalibrated model-assisted estimator (which subtracts the prediction directly) is highly sensitive to bias: when $|b|$ is on the order of the outcome SD, design effect climbs well above 1, i.e., adjustment \emph{hurts} precision. By contrast, the calibrated estimators absorb $b$ into the intercept during calibration, so their design effect is flat in $b$.

\paragraph{Comparison with AIPW and digital twin estimators.}
\label{app:sim-aipw}
To directly compare our calibrated regression adjustment against a semiparametric hybrid alternative, we add AIPW with an estimated propensity score to the simulation. We use $\hat{f}_0, \hat{f}_1$ as outcome models and logistic regression on $X$ for $\hat{e}(X_i)$. We deliberately estimate the propensity score rather than supplying the known design propensity, since with the known propensity AIPW coincides with the model-assisted estimator already considered above.
We run 500 simulations ($n=500$, balanced RCT, $R^2_m = 0.75$, $R^2_\robot = 0.25$) under two bias regimes. In the \emph{symmetric} regime both $\hat{f}_0$ and $\hat{f}_1$ receive the same additive shift $b$, so the bias cancels in any treatment-minus-control contrast. In the \emph{asymmetric} regime only $\hat{f}_0$ is shifted by $b$ while $\hat{f}_1$ is unbiased, representing a realistic setting where the AI model predicts control outcomes less accurately than treatment outcomes. 

\begin{table}[htbp]
\centering
\small

\begin{subtable}{\linewidth}
\centering
\caption{Symmetric bias ($b$ applied equally to $\hat{f}_0$ and $\hat{f}_1$).}
\label{tab:sim-bias-symmetric}
\resizebox{\linewidth}{!}{%
\begin{tabular}{lccccc}
\toprule
Estimator & $b=0$ & $b=2.1$ & $b=4.7$ & $b=7.4$ & $b=10$ \\
\midrule
Unadjusted                  & $-0.002$ / $0.952$ & $+0.003$ / $0.954$ & $+0.001$ / $0.964$ & $-0.001$ / $0.968$ & $-0.002$ / $0.970$ \\
Regression       & $-0.002$ / $0.960$ & $+0.002$ / $0.956$ & $+0.002$ / $0.960$ & $+0.000$ / $0.964$ & $-0.002$ / $0.978$ \\
Digital twin                & $-0.001$ / $0.938$ & $+0.002$ / $0.954$ & $+0.001$ / $0.946$ & $+0.001$ / $0.928$ & $-0.001$ / $0.942$ \\
AIPW (est.\ $\hat{e}(x)$)   & $-0.002$ / $0.962$ & $+0.001$ / $1.000$ & $+0.002$ / $1.000$ & $+0.002$ / $1.000$ & $-0.002$ / $1.000$ \\
\bottomrule
\end{tabular}%
}
\end{subtable}

\vspace{1em}

\begin{subtable}{\linewidth}
\centering
\caption{Asymmetric bias ($b$ applied to $\hat{f}_0$ only; $\hat{f}_1$ unbiased).}
\label{tab:sim-bias-asymmetric}
\resizebox{\linewidth}{!}{%
\begin{tabular}{lccccc}
\toprule
Estimator & $b=0$ & $b=2.1$ & $b=4.7$ & $b=7.4$ & $b=10$ \\
\midrule
Unadjusted                  & $+0.000$ / $0.966$ & $+0.001$ / $0.938$ & $+0.003$ / $0.968$ & $+0.001$ / $0.946$ & $-0.000$ / $0.960$ \\
Regression      & $+0.000$ / $0.950$ & $+0.002$ / $0.962$ & $+0.000$ / $0.972$ & $-0.000$ / $0.952$ & $-0.000$ / $0.972$ \\
Digital twin                & $+0.000$ / $0.938$ & $-0.376$ / $0.000$ & $-0.853$ / $0.000$ & $-1.327$ / $0.000$ & $-1.796$ / $0.000$ \\
AIPW (est.\ $\hat{e}(x)$)   & $-0.001$ / $0.948$ & $+0.003$ / $1.000$ & $-0.003$ / $1.000$ & $-0.002$ / $1.000$ & $-0.000$ / $1.000$ \\
\bottomrule
\end{tabular}%
}
\end{subtable}

\caption{Bias / coverage across 500 simulations under symmetric (a) and asymmetric (b) bias regimes.}
\label{tab:sim-bias}
\end{table}
Under symmetric bias, all estimators stay unbiased, but only regression calibration retains correct 95\% coverage. AIPW overcovers as its standard error inflates quadratically in $b$ (collapsing power), and the digital twin estimator mildly undercovers. Under asymmetric bias, regression again maintains near-zero bias and correct coverage by absorbing arm-specific prediction errors as separate coefficients. By contrast, the digital twin estimator fails catastrophically (bias $-1.80$ at $b=10$, coverage $0\%$) and AIPW stays unbiased but again overcovers severely.


\section{Twin-2K-500 Implementation Details}\label{app:twin-details}

\paragraph{Outcomes, Predictions, and Coding.}
The outcome is a binary indicator for choosing the target behavior under each framing. We map survey responses into this indicator following the Twin-2K-500 protocol and apply the same mapping to both human responses and model predictions so that correlations and regression adjustments are directly comparable.

For LLM predictions, we use GPT-5.4 with the same temperature setting as the original Twin-2K-500 Defaults configuration and a maximum output length of 100 tokens. Each participant's full persona profile, approximately 127,000 characters of text derived from their responses to 500 prior survey questions, is provided without truncation. The system prompt follows the Twin-2K-500 protocol: ``You are an AI assistant. Your task is to answer the `New Survey Question' as if you are the person described in the `Persona Profile'.'' We query the model separately for each treatment condition (opt-in and opt-out) to obtain both $\robot_i(0)$ and $\robot_i(1)$.

To avoid the cost of sampling multiple times (which would require $100{\times}$ more API calls), we extract continuous probability predictions directly from the model's output logprobs. For each persona-condition pair, we make a single API call with \texttt{logprobs=True} and \texttt{top\_logprobs=10}, then extract $P(\text{``1''})$ and $P(\text{``2''})$ from the returned token probabilities. We normalize these to obtain $P(\text{target}) = P(\text{``1''}) / (P(\text{``1''}) + P(\text{``2''}))$, or vice versa depending on which response indicates the target behavior. This yields a continuous prediction $\robot_i(z) \in [0, 1]$.

\paragraph{Feature Engineering, Selection, and ML Baselines.}
For traditional ML methods, we engineer 26 composite features from the raw survey data. Each composite is the arithmetic mean of all items belonging to that scale, identified by column name prefix in the Twin-2K-500 dataset (e.g., all columns matching \texttt{w1\_GREEN*} form the GREEN scale). The composites span four survey waves.

\emph{Wave 1} includes GREEN environmental attitudes (6 items), Empathy (20 items), Big Five personality (44 items mapped to 5 subscales by equal partitioning), Need for Cognition (18 items), Minimalism (12 items), Agentic-Communal orientation (24 items), plus demographics (region, sex). \emph{Wave 2} includes Beck Anxiety Inventory (21 items), Risk Aversion (3 subscales), Loss Aversion (4 subscales), Individualism (16 items), and Social Desirability (13 items). \emph{Wave 3} includes Need for Closure (15 items), Self-Monitoring (13 items), Need for Uniqueness (12 items), and Self-Concept Clarity (12 items). Missing values are imputed with the median. Features are then $z$-score standardized.

To avoid data leakage from data-driven feature selection, we use GPT-5.4 to select 10 theoretically relevant features for each outcome based on psychological theory. The LLM is prompted with descriptions of all 26 composites and the outcome definition and asked to select features most likely to predict the behavior. This selection is performed before seeing any outcome data. For \emph{Organ Donation}, the LLM selected empathy, agreeableness, agentic-communal orientation, risk aversion (health), loss aversion, individualism, social desirability, need for closure, region, and sex. For \emph{Green Energy}, it selected GREEN attitudes, minimalism, conscientiousness, openness, risk aversion, loss aversion, need for cognition, social desirability, region, and sex.

\paragraph{Cross-Fitting and Model Specifications.}\label{app:cross-fitting}
All ML models use a T-Learner approach with leave-one-out (LOO) cross-validation to generate out-of-sample predictions. We fit separate models for treatment and control arms. For each unit in the control group, we leave it out, fit on the remaining control units, and predict. We do the same for the treatment group. This yields within-arm predictions $\hat{f}_0(X_i)$ for control units and $\hat{f}_1(X_i)$ for treated units. The approach is deterministic, stratifies by treatment arm, and uses all $n-1$ observations for each prediction.

We use Ridge ($\alpha=1.0$) rather than Lasso for stable shrinkage at the small per-fold sample sizes in LOO cross-fitting, a shallow regularized XGBoost (depth-2, 25 trees), and default TabPFN. LOO cross-validation is deterministic.

\paragraph{Binary vs.\ Continuous LLM Predictions.}\label{app:prediction-methods}
We compare two approaches for obtaining LLM predictions using GPT-5.4. \emph{Binary} predictions come from a single deterministic call (temperature=0). \emph{Logprobs} predictions are continuous probabilities extracted from token logprobs (temperature=0). 
%
%
Logprobs predictions outperform binary in both experiments. For Organ Donation, logprobs achieves $\rho = 0.13$ (2.1\% VR) compared to $\rho = 0.11$ (1.6\% VR) for binary. For Green Energy, logprobs achieves $\rho = 0.37$ (13.7\% VR) versus $\rho = 0.36$ (13.0\% VR) for binary. The improvement is consistent with logprobs providing a softer, more calibrated probability than a hard binary prediction.

\section{Hillstrom Implementation Details}\label{app:hillstrom-details}

Each customer has 8 pre-treatment covariates. Recency measures months since last purchase. History\_Segment is a spending tier. History is dollars spent in the past year. Mens and Womens are binary indicators for prior merchandise purchases. Zip\_Code is Urban, Suburban, or Rural. Newbie indicates new customers. Channel is Phone, Web, or Multichannel.

Since the covariates are tabular, we convert them to text for LLM input using a structured key-value format:
\begin{verbatim}
Customer: recency=2 months, history=$150.00, 
history_segment=$100-$200, mens_buyer=yes, 
womens_buyer=no, zip=Suburban, newbie=no, channel=Web
\end{verbatim}

We tested four LLM prediction approaches. Binary classification (temperature=0) and logprobs extraction from token probabilities both produced poorly calibrated predictions in which the LLM was overconfident, predicting near-zero visit probability for most customers and yielding no variance reduction. Multi-draw sampling (temperature=1) showed similar issues.

Direct probability elicitation resolved this calibration problem. Instead of asking for a binary yes/no classification, we prompt the LLM to provide a numeric probability estimate:
\begin{verbatim}
What is the probability (0-100) this customer visits 
the website in the next 2 weeks? Think about how 
recency, purchase history, and channel affect visit 
likelihood. Reply with just a number between 0 and 100.
\end{verbatim}

This approach produces well-differentiated predictions that correlate with actual outcomes. We use GPT-5.4 with temperature=0 for determinism. Control predictions range from 8\% to 84\% with a mean of 33\%, while treatment predictions range from 8\% to 86\% with a mean of 42\%.

For ML comparison, we use the raw tabular features directly, one-hot encoding categoricals and standardizing numerics. Models include Ridge regression, XGBoost, and TabPFN, all with T-Learner LOO or 10-fold CV to generate out-of-sample predictions.

\section{Tech experiment implementation details}
\label{app:tech-experiment-details}

\paragraph{LLM-based user intent prediction pipeline.}
To construct auxiliary signals for variance reduction, we generate LLM predictions of user intent via a three-step pipeline that transforms structured user activity logs into human-readable summaries of recent engagement and applies reasoning models to infer latent conversion intent.
The three steps are as follows:
\begin{enumerate}
    \item \emph{Sessionization of User Activity.} 
    Raw timestamped user activity logs---including search interactions, browsing behavior, and conversion-related actions---are aggregated into fixed 15-minute sessions.
    These windows capture short-term behavioral dynamics while enabling analysis of how engagement patterns evolve over a user’s recent activity history.
    
    \item \emph{Text Narrative.} 
    Structured activity data from Step 1 are converted into a human-readable textual narrative using a deterministic templating specification that summarizes each user's most recent 7-day engagement history, broken down into time windows. The narrative includes basic user profile attributes and a chronological description of search, browsing, and conversion-related behaviors. This template functions analogously to a \textit{virtual judge prompt}, explicitly defining the evidence presented to the model, the behavioral signals to consider, and a fixed output schema.
    
    \item \emph{LLM-Based Intent Summary.} 
    The narratives from Step 2 are processed by o3, an LLM with explicit reasoning capabilities, instructed to analyze how user behavior evolves over time and to assess conversion intent. 
    The reasoning model (i) summarizes activity trends, (ii) assigns a discrete customer journey stage label (Exploring, Considering, Ready-to-Convert, or Other), and (iii) produces a continuous conversion likelihood in $\left[0,1\right]$ as a quantitative measure of conversion intent. The output also includes a structured explanation and confidence score for its prediction.
    We use these conversion likelihood scores as soft labels of latent user intent, as well as our primary LLM-based prediction signal.

\end{enumerate}

\paragraph{Fixed-horizon vs.\ sequential analysis.}
Our analysis targets a fixed cohort of users entering on the experiment's first day, with a single treatment effect estimate at the end of the month-long window---standard practice on many large A/B testing platforms.
We do not analyze time-to-significance, which would additionally depend on user entry timing, journey length, and seasonality.
Extending AI-assisted variance reduction to sequential or anytime-valid settings is a natural direction for future work.
\end{document}